\documentclass[sigconf, anonymous=false]{acmart}
\usepackage[utf8]{inputenc}
\usepackage{balance}
\usepackage{amsmath}
\usepackage{mathrsfs} 
\usepackage{bbm}
\usepackage{dsfont}
\usepackage{algorithm}
\usepackage{algorithmic}
\usepackage{colortbl}
\usepackage{tikz}
\usetikzlibrary{shapes,arrows,spy,positioning,decorations,decorations.pathreplacing}

\copyrightyear{2020}
\acmYear{2020}
\setcopyright{acmcopyright}\acmConference[UMAP '20]{Proceedings of the 28th ACM Conference on User Modeling, Adaptation and Personalization}{July 14--17, 2020}{Genoa, Italy}
\acmPrice{15.00}
\acmDOI{10.1145/3340631.3394846}
\acmISBN{978-1-4503-6861-2/20/07}

\title{Opportunistic Multi-aspect Fairness \\through Personalized Re-ranking}

\author{Nasim Sonboli}
\email{nasim.sonboli@colorado.edu}
\affiliation{%
  \institution{University of Colorado, Boulder}
  \city{Boulder}
  \state{Colorado}
  \country{USA}
}

\author{Farzad Eskandanian}
\email{feskanda@depaul.edu}
\affiliation{%
  \institution{DePaul University}
  \city{Chicago}
  \state{Illinois}
  \country{USA}
}
\author{Robin Burke}
\email{Robin.Burke@colorado.edu}
\affiliation{%
  \institution{University of Colorado, Boulder}
  \city{Boulder}
  \state{Colorado}
  \country{USA}
}

\author{Weiwen Liu}
\email{wwliu@cse.cuhk.edu.hk}
\affiliation{%
  \institution{The Chinese University of Hong Kong}
  \city{Shatin, Hong Kong}
  \country{China}
}

\author{Bamshad Mobasher}
\email{mobasher@cs.depaul.edu}
\affiliation{%
 \institution{DePaul University}
 \city{Chicago}
 \state{Illinois}
 \country{USA}}

\begin{document}
\fancyhead{}

\begin{abstract}
As recommender systems have become more widespread and moved into areas with greater social impact, such as employment and housing, researchers have begun to seek ways to ensure fairness in the results that such systems produce. This work has primarily focused on developing recommendation approaches in which fairness metrics are jointly optimized along with recommendation accuracy. However, the previous work had largely ignored how individual preferences may limit the ability of an algorithm to produce fair recommendations. Furthermore, with few exceptions, researchers have only considered scenarios in which fairness is measured relative to a single sensitive feature or attribute (such as race or gender). In this paper, we present a re-ranking approach to fairness-aware recommendation that learns individual preferences across multiple fairness dimensions and uses them to enhance provider fairness in recommendation results. Specifically, we show that our opportunistic and metric-agnostic approach achieves a better trade-off between accuracy and fairness than prior re-ranking approaches and does so across multiple fairness dimensions.

\end{abstract}

\maketitle

\section{Introduction}

Recommender systems are designed to assist users to find items of interest. Such systems model users' historical behaviors and generate personalized recommendations tailored to users' interests or needs. Recent research has identified a key limitation in a user-focused approach to recommender systems development, namely that it ignores multistakeholder aspects of the systems in which recommendation is embedded\cite{abdollahpourimulti2020}. In particular, the problem of \textit{provider fairness} has been underappreciated in recommender systems research, as it concerns the impact of recommendation delivery on the providers of items being recommended and the questions of fair treatment that may arise\cite{burke2017multisided}.

Recent research has sought to alleviate this concern using a variety of approaches. See, for example, \cite{yao2017beyond,burke2018balanced,ekstrand2018exploring,liu2019personalized,kamishima2016model,beutel2019fairness}. What these approaches share is that they focus on a single dimension over which fairness is sought: a single protected group among the providers, and except for \cite{liu2019personalized}, they do not take user preferences in item features into account. 

The problem of promoting provider fairness while maintaining recommendation accuracy can be generally characterized as a multi-objective optimization problem. If optimal fairness and optimal recommendation accuracy could be achieved simultaneously, there would be no need for research in this area. However, optimizing recommendation accuracy often comes at the expense of provider fairness, due to various biases present in recommender systems, including popularity bias \cite{celma2008hits,lee2014fairness}, and user-base composition \cite{lin2019crank, yao2017beyond}. Research in provider fairness is therefore generally concerned with improving the tradeoff between fairness and accuracy, or in other words, increasing the amount of fairness that can be gained for a given degree of accuracy loss.

Rather than look for improvements through global optimization as in \cite{yao2017beyond}, our work in this paper extends the approach pioneered in Liu, et al. \cite{liu2018personalizing,liu2019personalized} of seeking to improve the accuracy / fairness tradeoff through increased \textit{personalization}. Namely, can we tailor the type and degree of optimization specific to each user's tastes and preferences and therefore improve accuracy? We label this approach \textit{opportunistic} because we view each user as presenting a particular type of opportunity to increase recommendation fairness and try to make the most of each. In particular, we seek to identify the particular dimensions along which a user might be open to result diversification that improves fairness and thereby enable multiple fairness concerns to be addressed at once.

\begin{table}[htb]
    \begin{tabular}{|c|c|c|c|c|}
    \hline
        $user_{1}$ & $F_{1}$:Region & $F_{2}$:Gender & $F_{3}$:Sector & $F_{4}$:Amount \\
    \hline
        $item_{1}$ & Africa & Female & Agriculture & \$0-\$500\\
    \hline
        $item_{2}$ & Africa & Female & Health & \$0-\$500\\
    \hline
        $item_{3}$ & Africa & Female & Clothing & \$0-\$500 \\
    \hline
    \end{tabular}
    \caption{Profile of $user_{1}$}
    \label{table:user_profile}
\end{table}

As an example, in the context of loan recommendation, suppose user $u$ prefers to lend her money to women in Kenya but she does not have a strong preference for a loan's purpose or economic sector. This user's profile might appear as in Table \ref{table:user_profile}. While the user might not respond well to loans in other countries, we can consider her open-mindedness regarding the Sector feature as an opportunity to increase fairness in this area. For the sake of example, assume loans from the Education and Conflict Zones sectors are historically underfunded in Kenya, so the loans in these sectors are identified as protected. Consider the recommendation results in Table \ref{table_user_recoms}. The first two recommendations ($r_1 and r_2$) increase fairness across only the Sector feature by promoting items from underfunded sectors while honoring the user's preference to lend money to Kenyan women. On the other hand, loan $r_3$ might not be an effective recommendation for this user since it diversifies on the wrong dimensions, although it might still be promoting protected items. In other words, we want to promote fairness concerns when the user's profile indicates receptivity and be cautious otherwise.

\begin{table}[tbh]\centering
    \begin{tabular}{|c|c|c|c|c|}
    \hline
        $user_{1}$ & $F_{1}$:Region & $F_{2}$:Gender & $F_{3}$:Sector & $F_{4}$:Amount \\
    \hline
        $r_{1}$ & Africa & Female & Conflict Zones & \$0-\$500\\
    \hline
        $r_{2}$ & Africa & Female & Education & \$0-\$500\\
    \hline \rowcolor[gray]{0.7}
        $r_{3}$ & Asia & Male & Livestock & \$500-\$700\\
    \hline
    \end{tabular}
    \caption{Recommendations for $user_{1}$}
    \label{table_user_recoms}
\end{table}



This paper addresses the following research questions:
\begin{description}
    \item [RQ1:] Do users exhibit different patterns of preference across fairness dimensions?
    \item [RQ2:] Can these patterns be exploited to improve the recommendation fairness / accuracy tradeoff using re-ranking?
\end{description}

\section{Background}
\label{sec:background}
This line of research has much in common with work that seeks to enhance diversity in recommendation \cite{zhang2008MonotonyDiv, vargas2011rankRelDiv, mmr, eskandanian2017clusteringDiv}. However, the key differences have to do with the concerns being addressed and, accordingly, the way in which success is measured. Usually when diversity is invoked as a desirable property of a recommender system, it is in the service of some user-oriented goal. Diverse recommendations can help a system cope with a diverse range of user intents and contexts. For example, a restaurant recommender might know that a user sometimes goes to family-style pizzerias 70\% of the time and fancy French restaurants 30\% of the time. Rather than present just pizzerias in a recommendation list, even though that is likely to be the right answer statistically, it might be better to include one or two fine dining establishments on the list, just in case the user is looking for a ``date night'' recommendation this time around. 

Typical measures of diversity such as intra-list distance, for example \cite{ziegler2005improving}, therefore measure the difference among items in each user's list, without regard to what items they are. Diversity as a fairness concern seeks varied outputs for a completely different reason, namely to increase the prevalence of items from under-represented providers, and measures outcomes relatively to those providers specifically. We will distinguish between these sense of diversity by using the term \textit{list diversity} to refer to the user-centered objective and \textit{fairness-promoting diversity} to the provider-centered objective, our main concern in this paper.

Another related definition of diversity is what is called \textit{aggregate diversity} or catalog coverage. The question here is whether the recommender is presenting all of the available items in the catalog. This can be seen as a minimal form of fairness where the frequency of appearance is not considered, just that an item is recommended at least once, and we do not differentiate between different items or different providers~\cite{adomavicius2011improving}.

As noted above, most work in recommendation fairness, and machine learning fairness more generally, simplifies the problem of fairness-enhancement by concentrating on a single (usually binary) distinction between a protected group and an unprotected group. This is an excellent starting point and admits of tractable mathematical formulations. However, this approach is not a good match to real-world applications, where there are likely to be multiple fairness concerns related to multiple dimensions of identity \cite{kearns2019empirical}. 



\section{Problem formulation}

Given a set of users $\mathcal U=\{u_1,\ldots,u_n\}$, a set of items $\mathcal V=\{v_1,\ldots,v_m\}$, and initial ranking lists $R(u)$ for users $u\in \mathcal U$, our task is to re-rank $R(u)$ and generate a list of $k$ distinct items $S(u)$ that is both accurate and fair similar to \cite{liu2019personalized}'s goal.

We will further assume that each item $v_i\in\mathcal V$ is represented by a $d$-dimensional feature vector $\vec{\phi}_i = \langle f_{i1},\ldots,f_{id}\rangle$ over a set of categorical features $F = \{F_1, F_2, \ldots, F_d\}$. Each dimension $F_j$ can be viewed as a set of categorical values or labels and so for an item $v_i$, its feature vector $\phi_i$ contains $f_{ij} \in F_j$ for each feature $F_j$. We will use the notation $c_j = |F_j|$ to refer to the cardinality of the feature $F_{j}$.

As an example, suppose that our set of items are loans and users are our potential lenders. Suppose that each loan is characterized by two features: geographical region and economic sector. Thus, $F = \{\mathrm{Region}, \mathrm{Sector} \}$, and $d = 2$. Suppose that we have 5 geographical regions and 7 economic sectors. For example:  $\mathrm{Region} = F_1 = \{\mbox{Africa}, \mbox{Asia}, \mbox{Americas}, \ldots\}$ and $\mathrm{Sector} = F_2 = \{\mathrm{Agriculture}, \mathrm{Housing}, \mathrm{Education}, \mathrm{Conflict Zones}, \ldots\}$. If a particular loan $v_i$ is sought in the agriculture sector in Africa, we would say $\vec{\phi}_i = \langle \mathrm{Africa}, \mathrm{Agriculture} \rangle = \langle f_{i1}, f_{i2} \rangle$.



A protected class, within some $F_j$ feature, consists of a set of values $F_j' \subset F_j$ that are considered protected and for which fairness is sought. There may be multiple fairness dimensions of concern, we define the protected dimensions  $F'$ as the subset of $F$ that contain such protected values. For example, if Education and Conflict Zone loans are relatively underfunded, then in the Sector feature, these two specific values form the protected group $F'$. 
\subsection{Personalized diversity}

Studies have shown that users generally prefer more recommendation results they perceive as diverse \cite{hu2011enhancing}. This suggests that the opportunity for fairness-enhancing diversification exists and may come at minimal cost in terms of user experience. However, users differ in the variety that they seek in recommendations~\cite{tintarev2013adapting}. Some recommendation research has sought to capitalize on these differences in improving diversity \cite{eskandanianuser_2016}. Here we aim to do the same in a more fine-grained way, consider each user's interest in diversity across multiple features.

\begin{figure}[tbh!]
    \centering
    \includegraphics[width=8cm]{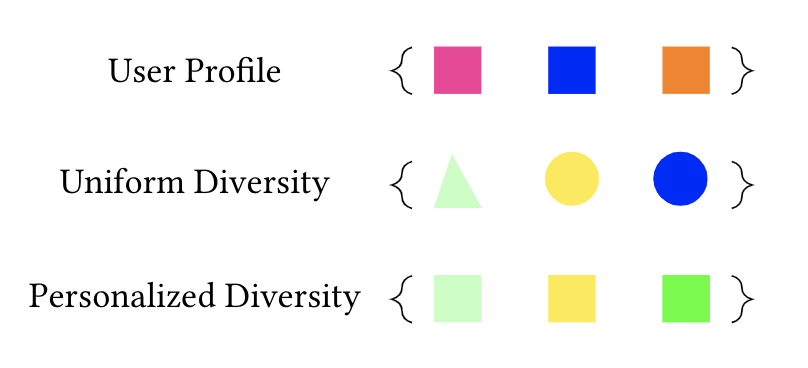}
    \caption{Uniform vs Personalized Diversity}
    \label{fig:uniform_vs_personalized_div}
\end{figure}

Figure \ref{fig:uniform_vs_personalized_div} gives a schematic depiction of this distinction. In this example, each item has a color and a shape feature. A user profile, shown at the top, consists of squares of different colors. Clearly, this user has a strong interest in squares and cares less about what color they are. A recommender that prioritizes triangles and circles as a protected group as well as greenish/yellowish hues might deliver recommendations as shown in the second row. These will likely not be accepted as they deviate too much from the characteristics preferred by the user. A better approach would be to diversify only in (the dimensions/values of) color, retaining the aspect of the items that the user apparently prefers.

Liu et al. \cite{liu2018personalizing,liu2019personalized} introduced the concept of recommendation re-ranking using a quantity $\tau_u$, a user-specific measure of interest in diversity, based on information entropy. Here we extend this definition to take into account multiple item features while seeking fairness within each feature's dimensions. Instead of a single user-specific $\tau_u$, the $\vec{\tau}_u$ vector will represent the user's level of tolerance for diversity across the feature space (such as the user in the above example having more tolerance for diversity in the ``color'' feature and less in the ``shape'' feature). Specifically,

\begin{equation}
\vec{\tau}_u(F_{j})\buildrel\triangle\over=-\sum_{f \in F_j} P(f|u)\log P(f|u),
\label{eq:tau}
\end{equation}
where $P(f|u)$ is computed as the fraction of items in the user's profile that have the feature value $f$. This can be interpreted as the user's likelihood of liking items with that value. The higher the entropy value is for a user on a feature, the higher her tolerance to see diversity within that feature. For example, the user in Table~\ref{table:user_profile} would have low entropy for Region and Gender, but higher entropy for Sector.

This vector of values, therefore, quantifies the relative opportunities for providing diverse results to users. As we show in Section 5.4, these values vary widely across different features and different users, motivating a recommendation techniques that is sensitive to these individual differences. 

\subsection{Recommendation re-ranking}

Re-ranking is a common technique for enhancing the non-accuracy properties of recommender systems output. It provides a relatively simple framework for augmenting an existing recommender system with concerns that are not part of its design. Generally speaking, a re-ranker is a function that maps a ranked list $R(u)$ of size $k$ (e.g., a ranked recommendation list) and produces a new list $S(u)$ of size $k'$ where $k'<= k$ and where all items are drawn from the original list: $\forall{i}: i \in S(u) ~\mathit{iff} i\in R(u)$. The loss of ranking accuracy in doing so is thereby limited by the size $k$; no item in $S(u)$ can be worse than what the original recommendation placed at rank $k$. 

Re-ranking algorithms of this type were introduced in information retrieval for enhancing user-oriented diversity. The \textit{Maximum Marginal Relevance} method as proposed in \cite{carbonell1998use} measures for each user, the dissimilarity between a query and the items in her retrieved results. This method intends to combine query relevance and list diversity using a greedy list accumulation algorithm. The algorithm builds the output list $S$ one item at a time. 

At each point in time, it scores potential new items by a combination of their relevance (as computed in the initial retrieval step) and their differences from the current list (novelty), computed by identifying the item $j \in S$ that is most similar to the new item.

In our context, we will assume that we have some function \textit{sim} that computes similarity between two items $i, j$ and that our recommender system returns a relevance score of \textit{rec(v,u)} for a user $u$ and item $v$. We can then define the MMR scoring function:


\begin{equation}
    \mathit{MMR}(u,v,R,S)\buildrel\triangle\over= \arg\max_{v \in R \setminus S} [\lambda  (\mathit{rec}(v,u) - (1 - \lambda) \sum_{v' \in S} \mathit{sim}(v,v')]
\label{eq:mmr}
\end{equation}

Effectively, the algorithm, at each point, finds the next item to include by incorporating the original ranking (as encapsulated in the recommendation score), but penalizes that score when the proposed item is highly similar to the items already added.

There is a subtle difference between the MMR formulation here and its original specification. When scoring a new item to decide whether to add it to the re-ranked list, MMR chooses the most similar item -- this is the ``marginal'' part of the algorithm. Our formulation calculates the summation of similarities between the target item and all the other items in the re-ranked list. We can think of this as identifying the item with maximum aggregate difference from the existing list. We will explain later how this change is appropriate in a fairness context.

\textit{eXplicit Query Aspect Diversification} method proposes another formulation to enhance diversity. Although, this method has a similar goal to MMR, it enhances diversity with respect to specific aspects of an item \cite{santos2015search}. The diversity objective relative to a particular aspect (e.g., feature, topic, or category) is considered satisfied if one item containing that aspect is added to the result list. In context of recommendations, we can express this ranking score as follows:

\begin{equation}
    \mathit{xQuAD}(u,v,R,S)\buildrel\triangle\over= \arg\max_{v \in R \setminus S} [\lambda  (\mathit{rec}(v,u) + (1 - \lambda) \max_{v' \in S} \mathds{1}_{\vec{v} \cap \Vec{v'} = \emptyset}],
\label{eq:xquad}
\end{equation}
\noindent where $x_v$ represents the set of aspects present in item $v$. In effect, this algorithm boosts the rank of items that, when added to the list so far, bring in new aspects -- features that have not yet appeared in the list. 

Liu et al. \cite{liu2018personalizing,liu2019personalized} proposed two extensions to xQuAD. The first \textit{FAR} (Fairness-Aware Reranking) applied the formalism using aspects of an item defined over a fairness-relevant feature. In this configuration, the algorithm boosts the scores of items from protected groups when no such item has yet been added to the list. Once the group is represented, the boosting disappears. This can be seen as an implementation of the ``Rooney rule'' \cite{kleinberg2018selection} that ensures minimum representation for protected groups. The second variant \textit{PFAR} adds personalization to this process. Using the $\tau_u$ information entropy measure described above, the fairness-boosting term is modulated so that users with more diverse profiles (who have a high diversity tolerance/higher entropy) are presented with results containing more fairness-enhancing diversity.

In particular, the scoring function of PFAR is composed of a personalization score $\mathit{rec}(v,u)$ and a personalized fairness score. PFAR simply assumes only one sensitive feature need to be considered. Suppose the given sensitive feature dimension is $F_a$, then the scoring function is defined by

\begin{equation}
\arg\max_{v \in R \setminus S}[\lambda\mathit{rec}(v,u) + (1-\lambda)\tau_u \min_{v' \in S} \mathds{1}_{v_{a} \neq v'_{a}}],
\end{equation}

where $v_{a}$ is the a-th element of the feature vector $\vec{v}$.
Note that PFAR inherits the limitation of xQuAD that it assumes binary inclusion as a sufficient definition of fairness and it is therefore difficult to tune it to improve the representation of protected groups in a proportional way. 


\section{Opportunistic fairness}
We are now ready to describe \textit{OFAiR} (\textbf{O}pportunistic \textbf{F}airness-\textbf{A}ware \textbf{R}eranking), which incorporates personalization at the feature level into the re-ranking process and also allows fine-grained control of protected group promotion by using per-feature weights. 

As discussed above, we can represent the variation in a user's profile across all features through the vector $\vec{\tau}_u$, calculated using information entropy. However, because these weights are feature-specific, we cannot incorporate them as a single multiplier as found in PFAR. Also, because we are interested in fine-grained control over the proportions of protected group items in recommendation lists, the xQuAD formula with its binary inclusion metric is not appropriate. So, our alternative in OFAiR applies the MMR approach by penalizing item similarity, but we build the feature significance into the similarity metric itself. We want to add items to the recommendation list if they add to the representation of protected groups in the recommendation list and if they differ from the items on the list in areas of high diversity tolerance for the user. To achieve this effect, we multiply together the user-specific tolerance weight for each feature and a weight associated with a feature's protected / unprotected class. 

We use weighted cosine similarity to allow the similarity between two items to be controlled by weights associated with each dimension. Because the weights actually vary by value, not just by dimension, and we can only pass a single weight vector to the weighted cosine similarity function, we convert the feature vector $\vec{\phi}$ to a smoothed binary vector of dummy variables $b_i$ with one dimension for each possible feature value. The smoothing operation means that instead of missing values being represented by zero, they have a small value $\epsilon = 2.2e^{-16}$. The user tolerance weights are correspondingly expanded in dimension to match: $\vec{\tau}_u \rightarrow \vec{\gamma}_u$. 

Let $\vec{a} \circ \vec{b}$ represent the element-wise (Hadamard) product between two vectors $a$ and $b$. Let $W(f')$ be a function that returns the weight of a particular binary feature value $f'$. This value will be small for unprotected values and larger for protected values as described below. For all items, we derive a weight vector $\vec{w}$ where the elements $w_j = W(f'_{j})$. Let $\vec{z_{u}}$ be the product, which combines the two types of weights. 
\begin{equation}
\vec{z}_u = \vec{\gamma}_u \circ W(F')
\end{equation}
The entries $z_{uj}$ represent the weight assigned to user $u$ for the $j$th dummy (smoothed binary) feature, combining both individual diversity tolerance and the system's fairness objective. 



The weighted cosine metric applies weights to the terms of the cosine computation:
\begin{equation}
    \mathit{wcos}(\vec{b}, \vec{b}', z_u)\buildrel\triangle\over= \sum_{j}^{|F|}{z_{uj} b_j \times b'_j} \frac{1} {\sqrt{\sum_{j} {z_{uj} b^2_j}} \times \sqrt{\sum_{j}{z_{uj} 
    b'^2_j}}}
\end{equation}


Two items are similar under this calculation if their values on many dimensions are the same and those dimensions are ones where the user profile has high entropy / variation and where their associated weight is high. 

Recall that the similarity calculation in MMR is used to penalize items that would be redundant with what is already in the recommendation list. So, the higher the similarities are, the higher the penalty. Therefore, we will want a weighting scheme where protected items are weighted high: their similarity is more important to the system.

This weighting scheme interacts with our aggregate difference alteration of the MMR algorithm noted above. By definition, protected items will be a small subset of the recommended items. Therefore, protected items will always differ from the list in aggregate. Also, the features in the recommendation list are likely to reproduce the consistencies in the user profile that represent lower tolerance for diversity. Weighting the protected features more highly helps promote diversity on those dimensions while keeping the other dimensions less diverse.

Various schemes for the weighting function were considered in our experimentation. In this paper, we report on a simple scheme where protected features receive a fixed high weight $\alpha$ and unprotected features a fixed low weight $\alpha/100$. In our experiments, the results were not sensitive to the magnitude of these values as long as protected features have a lower weighting. Additional exploration of feature weighting will be considered in future work.

\section{Experiments}\label{sect:exp}
\subsection{Evaluation Metrics}
The accuracy of the following methods was evaluated based on Precision, Recall, normalized discounted cumulative gain (nDCG), and to calculate their feature-based diversity both intra-list distance (ILD) and entropy of the recommendation lists were used. The fairness of lists was evaluated based on protected group exposure, which measures the fraction of the recommendation list that consists of protected group items. This value is related to the fairness concept of ``statistical parity,'' measured relative to items' level of promotion within the recommender system. Because list lengths are fixed (10 in our case), the exposure of unprotected items is just one minus the protected group exposure.

\subsection{Dataset}
We test our model on two datasets. The first is The Movies Dataset, which was obtained from the Kaggle website and contains the metadata of 45,000 movies listed in the Full MovieLens Dataset \footnote{https://grouplens.org/datasets/movielens} which were released on or before July 2017. Although movies are not a domain to which important fairness concerns are typically applied, we use this dataset as a well-known example with a rich set of provider-side features. The dataset contains 26 million ratings from 270,000 users for all 45,000 movies. Ratings are on a scale of 1-5. Each movie contains a set of features from which the following were used in this project: genres, original language, release date, revenue, run-time, popularity, production countries and spoken language. A sample of this dataset was extracted which contained the 559,070 ratings from 6,000 users on 14,623 items (density of 0.63\%).

All the features were transformed into categorical variables. If the movie's popularity is greater than the average popularity, we tag the movie as popular and unpopular otherwise. We transform the revenue and run-time in the same way as well. The release date is bucketed into old and new if the movie's release date is before or after 1990 \cite{kamishima2016model}. All the categorical features were transformed into dummy variables, resulting in a total of 323 binary features.

For the purposes of exposition, we selected two features in each dataset along which to identify protected features, although the OFAiR algorithm supports any number of sensitive features. In the Movies Dataset, we identified the following protected classes within each feature: ``unpopular'' (popularity), ``lower revenue'' (revenue) , ``longer'' (running time), ``before 1990'' (release date), some genres and movies the were produced in some non-US countries. More specifically, in our experiments, within genre and production country features we chose ``Horror'', ``Music'', ``Mystery'', ``History'' (genres) and ``CA'', ``ES'', ``DE'', ``HK'' (countries) to be the protected group. These feature values were chosen because they represented a minority within each feature, and so are good exemplars for demonstrating the capabilities of our algorithm.

Our algorithms are also evaluated on a proprietary dataset obtained from Kiva.org, including all lending transactions over an 12-month period. Initially, there were 1,084,521 transactions involving 122,464 loans and 207,875 Kiva users. Of these loans, we found that 116,650 were funded, that is they received their full funding amount from Kiva users by the 30-day deadline imposed by the site. We selected only the funded loans for analysis. Each loan is specified by features including borrower's name/id, gender, borrower's country, loan purpose, funded date, posted date, loan amount, loan sector, and geographical coordinates. 
To reduce the feature space, and to solve the multicollinearity problem, highly correlated features were removed. The percentage funding rate (PFR) was added as a new feature, computed as follows:
\begin{equation}
 \mbox{\textit{PFR}} =  \frac{1}{\mbox{\textit{{\#} days to fund}}} * 100 
\end{equation}

The percentage funding rate captures the speed at which a loan goes from being introduced in the system to being fully funded.\footnote{Loans not fully funded within 30 days are dropped from the system and the money raised is returned to lenders.} For example, a loan with PFR of 25\% is accumulating a quarter of its needed capital each day. After preparing the data, the final features for each loan reduced to borrower's gender, borrower's country, loan purpose, loan amount (binned to 10 equal-sized buckets), and loan's percentage funding rate. We found that this dataset was highly sparse (density = $4.2e^{-5}$) and could not support effective collaborative recommendation, because a loan can only attract a limited amount of support (up to that needed for its funding). There are no ``blockbuster'' loans with thousands of lenders.

To generate a denser dataset with greater potential for user profile overlap, we applied a content-based technique creating \textit{pseudo-items} that represent groups of items with shared features. We applied agglomerative hierarchical clustering \cite{rokach2005clustering} using the features of borrower gender, borrower country, loan purpose, loan amount (binned to 10 equal-sized buckets), and percentage funding rate (4 equal-sized buckets). We chose the cluster with the highest Silhouette Coefficient \cite{rousseeuw1987silhouettes} of around 0.69 which indicates a reasonable cohesion of the clusters. Then we applied a 10-core transformation, selecting pseudo-items with at least 10 lenders who had funded at least 10 pseudo-items. The retained dataset has 2,673 pseudo-items, 4,005 lenders and 110,371 ratings / lending actions.

In this dataset, we observed an imbalance within the following feature values/dimensions: (percentage funding rate), (country), (economic sector), (loan amount), (borrower gender). In keeping with Kiva's mission of providing equal access to capital across regions and economic sectors, we designate the items from the sectors and countries that have less than 1\% frequency in the training data as the protected group. More specifically 5 loan purposes in the economic sectors and 23 countries were selected to be the protected group. 
Although in both datasets we chose two features to achieve fairness within their multiple dimensions, our method supports choosing any number of such features.

\subsection{Variation in diversity tolerance}
\begin{figure}[tbh]
    \centering
    \includegraphics[width=\linewidth]{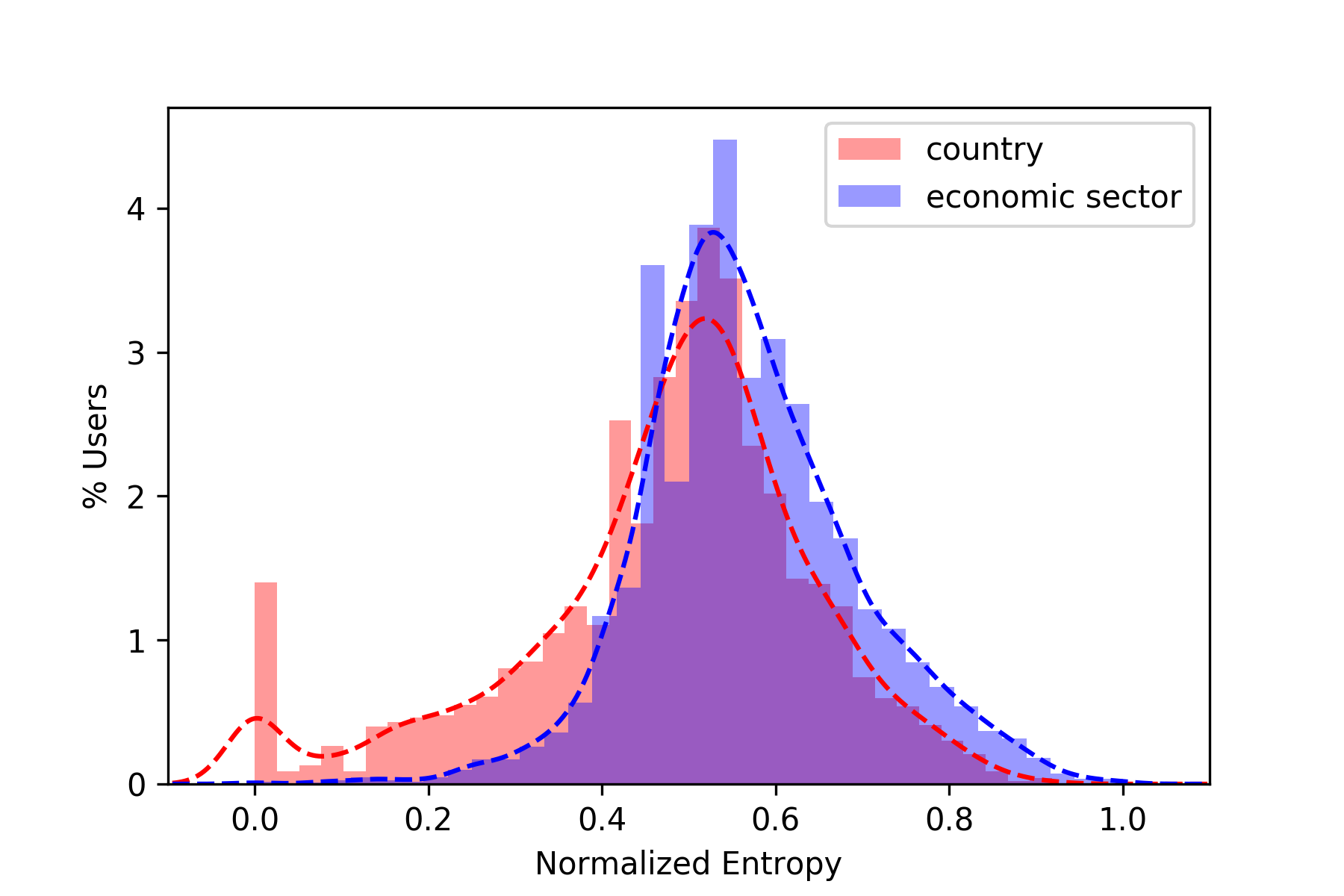}
    \caption{User tolerance value ($\tau$) for Economic Sector and Loan Country features in Kiva dataset.}
    \label{fig:tau_vector_Kiva}
\end{figure}

By examining the $\vec{\tau}$ vectors for each user, we can get evidence for RQ1: Do users exhibit different patterns of preference across fairness dimensions?
Figure~\ref{fig:tau_vector_Kiva} shows the $\tau$ values computed across for all users in the Kiva dataset. As the figure shows, users differ significantly in their profile entropies as measured for features of country and economic sector. (The differences across features are not meaningful, as they are a function of the prevalence of different feature values.) Some users have loans that vary widely across different economic sectors (shown in blue); others less so. Similar variety can be seen in country as well (shown in red), including some users who have loaned only to a single country.


\begin{figure}[tbh]
    \centering
    \includegraphics[width=\linewidth]{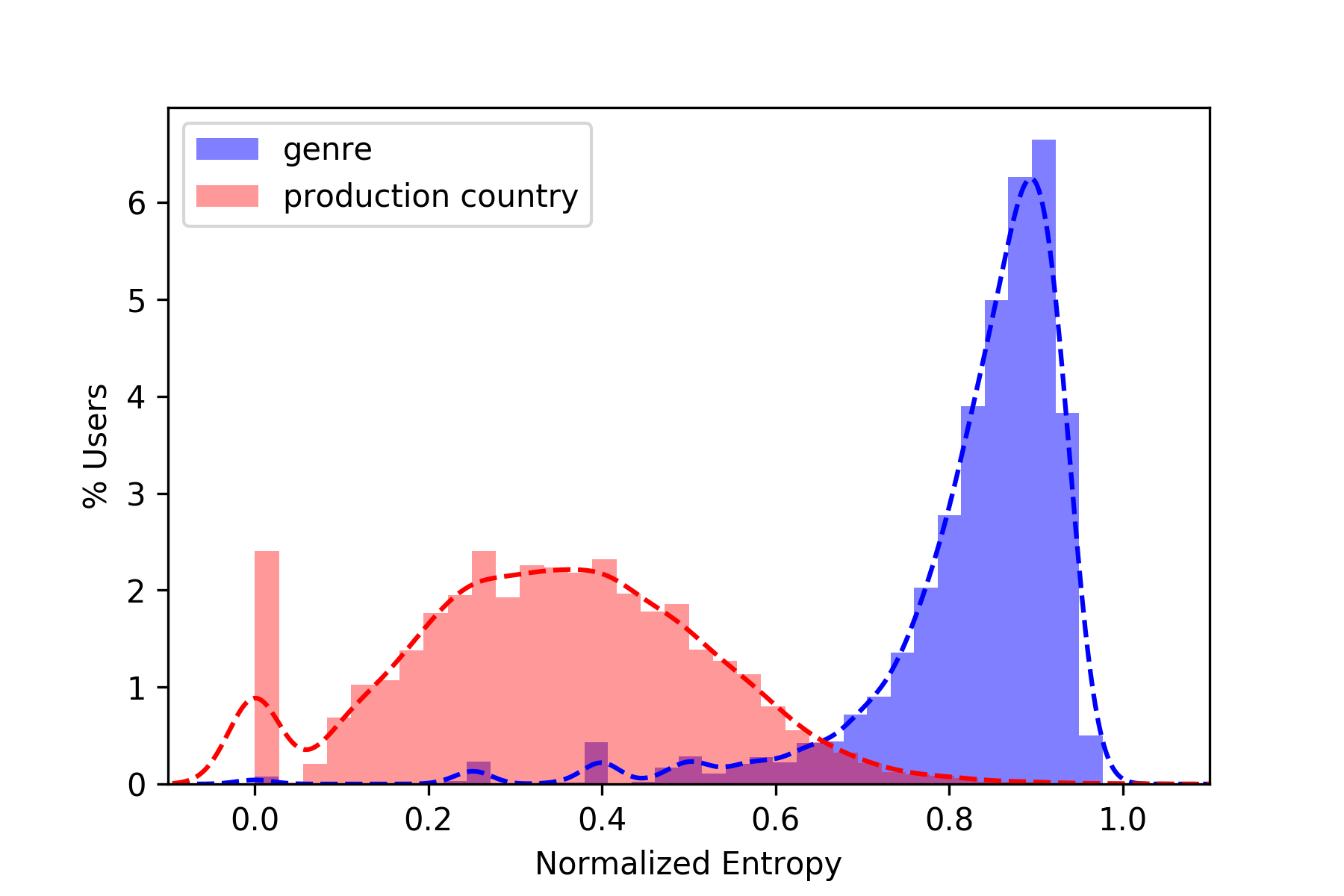}
    \caption{User tolerance value ($\tau$) for Genre and Production Country features in The Movies dataset.}
    \label{fig:tau_vector_ML}
\end{figure}

Figure~\ref{fig:tau_vector_ML} shows similar results for the Movies dataset. Again, we see that users in this sample have wide individual variance in the computed $\tau$ values for different dimensions of movies. For example, the variation in the entropy for the genre dimension (shown in blue) indicates that most of the users are watching movies from various genres while there are some users who usually prefer to watch the same few genres. The variation in the production countries (shown in red) is flatter and farther to the left, indicating users' narrower choice of movies in this dimension. Possibly, these viewers mostly watch movies that are produced in their countries or in their language.

We note that different features have different baseline entropy values in each dataset. In our future work, we plan to explore a refinement of the personalized tolerance measure using conditional entropy to calculate how much each user profile adds or detracts from the entropy in a particular feature.

\subsection{Comparing re-ranking algorithms}
We use non-negative matrix factorization as our baseline recommendation component. The algorithm was tuned on each dataset separately to achieve the best nDCG. The algorithm was trained on 80\% of the data and tested on the remaining 20\%. The nDCG of NMF was around 0.11 on the ML dataset and 0.076 on the Kiva dataset.
For each algorithm, we retrieve $k=200$ top items for each user and re-rank the list retaining the top $k'=10$ items.  

In our experiments, we compared our OFAiR algorithm with FAR and PFAR, as our baseline methods. We also used MMR by itself, as a diversity-enhancing re-ranker, a variant of OFAiR that includes only user tolerance weights for each feature, and a variant that includes only the fairness weights for the protected feature dimensions without the tolerance weights. In this way, we can study separately the contribution of each of these aspects of the algorithm.

\begin{table}[]
\begin{tabular}{lllll}
 Algorithm &  1\% & 2\% & 3\% \\
 \hline
 
 FAR & 12.06\% & 12.09\% & 12.12\% \\
 PFAR & 12.07\% & 12.08\% & 12.09\% \\
 MMR &  12.22\% & 12.67\% & 13.08\% \\
 MMR w/ tolerance & 12.83\% & 13.29\% & 12.66\%   \\
 MMR w/ fairness & 14.0\% & 15.14\% & 17.03\%  \\
 OFAiR & 16.76\% & 20.14\% & 22.81\% \\
 \hline
\end{tabular}
\caption{Fairness vs \% Accuracy Loss. Kiva dataset. 
Larger values mean improved fairness at the given accuracy level.}
\label{tbl:kiva_fairness_accuracy_relationship}
\end{table}


Table \ref{tbl:kiva_fairness_accuracy_relationship} summarizes the results across the different algorithms. We indicate the tradeoff between fairness and accuracy by reporting the (interpolated) protected item exposure at different levels of nDCG loss: 1\%, 2\% and 3\%. We arrive at the exposure values in the table by assuming a locally-linear relationship of nDCG and fairness/exposure in between different $\lambda$ values, basically locating intercepts in the tradeoff graph. (See below.) The table shows that FAR and PFAR do little to improve fairness in this setting. This is not surprising as these algorithms were designed for a situation in which fairness across a number of different providers is sought, rather than the protected item balance situation here. In Figures \ref{fig:kiva_mmr_based_methods}, and \ref{fig:ML_mmr_based_methods} below, we will omit FAR and PFAR for this reason. Of the other algorithms, we see a small advantage for OFAiR at the 1\% level of loss, increasing greatly at higher levels of loss. Both tolerance weights and fairness weights contribute to the results but their synergy in the OFAiR algorithm is apparent. It must be noted that in absolute terms, the fairness enhancement is somewhat disappointing. 16.76\% to 20.14\% increase still means that only 1.2 protected items will appear (on average) in each user's recommendation list. 

Table \ref{tbl:ML_fairness_accuracy_relationship} shows even stronger findings in favor of the OFAiR algorithm on the Movies dataset. Two trends are noticeable. One is that there is very little change in fairness for increased $\lambda$ values in the MMR and MMR with tolerance cases. This trend also exists in Kiva dataset. OFAiR is a clear improvement at all levels of nDCG loss, although in absolute terms the improvement is still small.

\begin{table}[]
\begin{tabular}{lllll}
 
 Algorithm &  1\% & 2\% & 3\% \\
 \hline
 
 FAR & 28.65\% & 28.64\% & 28.64\% \\
 PFAR & 28.63\% & 28.63\% & 28.63\% \\
 MMR &  28.44\% & 29.28\% & 29.92\% \\
 MMR w/ tolerance & 28.99\% & 30.83\% & 32.13\%  \\
 MMR w/ fairness & 32.51\% & 34.44\% & 35.85\%  \\
 OFAiR & 36.59\% & 39.41\% & 41.34\% \\
 
 \hline
\end{tabular}
\caption{Fairness vs \% Accuracy Loss. The Movies Dataset.}
\label{tbl:ML_fairness_accuracy_relationship}
\end{table}

Figure \ref{fig:kiva_mmr_based_methods} shows the results on the Kiva dataset for just the MMR-based algorithms: MMR, OFiAR, and the two versions incorporating different aspects of the OFAiR algorithm, tolerance weights (users) only, and fairness weights (items) only. The figure compares ranking accuracy in the form of nDCG versus the average exposure for protected items across recommendation lists. The figure gives a more complete picture of this tradeoff than the tables above, but generally tells the same story.

The general trend shows that by incorporating re-ranking, the algorithms move the fraction of protected group items from around 11\% to greater than 34\%. At the higher values of $\lambda$, the algorithms are quite similar, as might be expected. When we push the algorithms to focus more on fairness, differences emerge. The OFAiR and the MMR variant with only fairness weights are very similar until we get to nDCG loss around 0.1\%. At this point, the OFAiR algorithm dominates this tradeoff in terms of nDCG while keeping the fairness comparable. MMR and MMR with tolerance have curves that are essentially vertical, with very small fairness gain from diversification.

\begin{figure}[tbh]
    \centering
    \includegraphics[width=\linewidth]{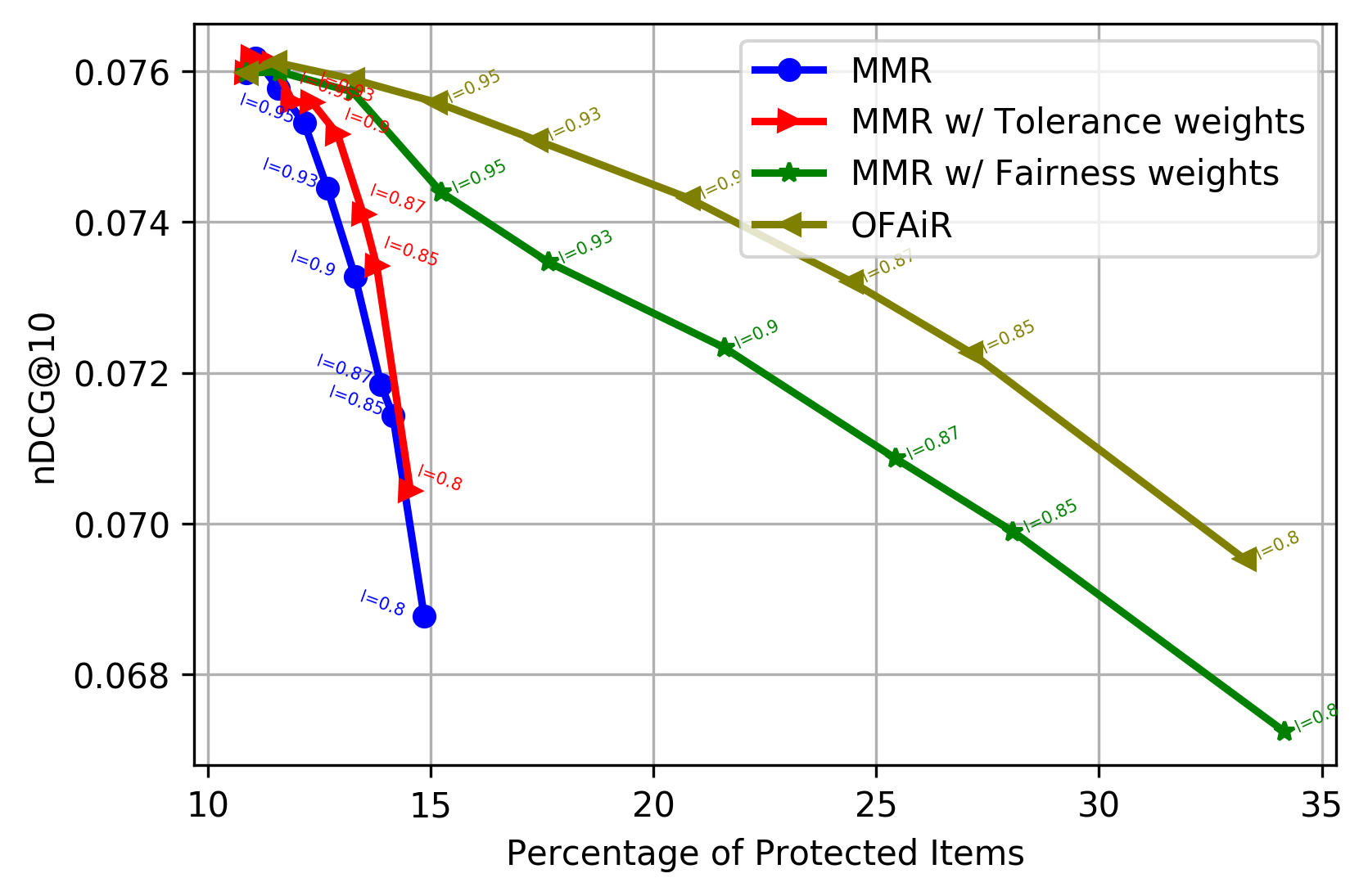}
    \caption{MMR-based re-ranking methods. Kiva dataset.}
    \label{fig:kiva_mmr_based_methods}
\end{figure}

Figure~\ref{fig:ML_mmr_based_methods} shows similar results for the Movies dataset. As suggested by Table \ref{tbl:ML_fairness_accuracy_relationship}, both MMR and MMR with tolerance fare poorly as fairness is emphasized. \footnote{Although note the small but intriguing bump for the tolerance-weight-based algorithm near $\lambda = 0.95$).}
This finding highlights the difference between a user-centered view of diversification, which MMR is targeted towards, and a fairness-oriented, provider-centered view. This effect may be due to the large feature diversity present in the Movies dataset. There are many ways for movies to be diverse without falling into the protected group. 

The difference between datasets is also apparent in the relative performance of the tolerance-weighted and the feature-weighted version of the algorithm. In the Kiva dataset, fairness weights greatly enhanced fairness, competing with the OFAiR algorithm at some points in the parameter space while in the Movies dataset OFAiR surpasses all the others except in higher lambdas. The other difference is in the effect of these algorithms on the percentage of protected items achieved. As it is shown, we achieve higher fairness gains in the Movies compared to the Kiva dataset.
These differences in performance could be due to domain differences in feature distributions, such that diversification along a preferred dimension does not necessarily yield protected items. The feature weights are needed to shift the algorithm's attention to the protected parts of the feature space. As before, much larger fairness gains are possible with OFAiR.


\begin{figure}[tbh]
    \centering
    \includegraphics[width=\linewidth]{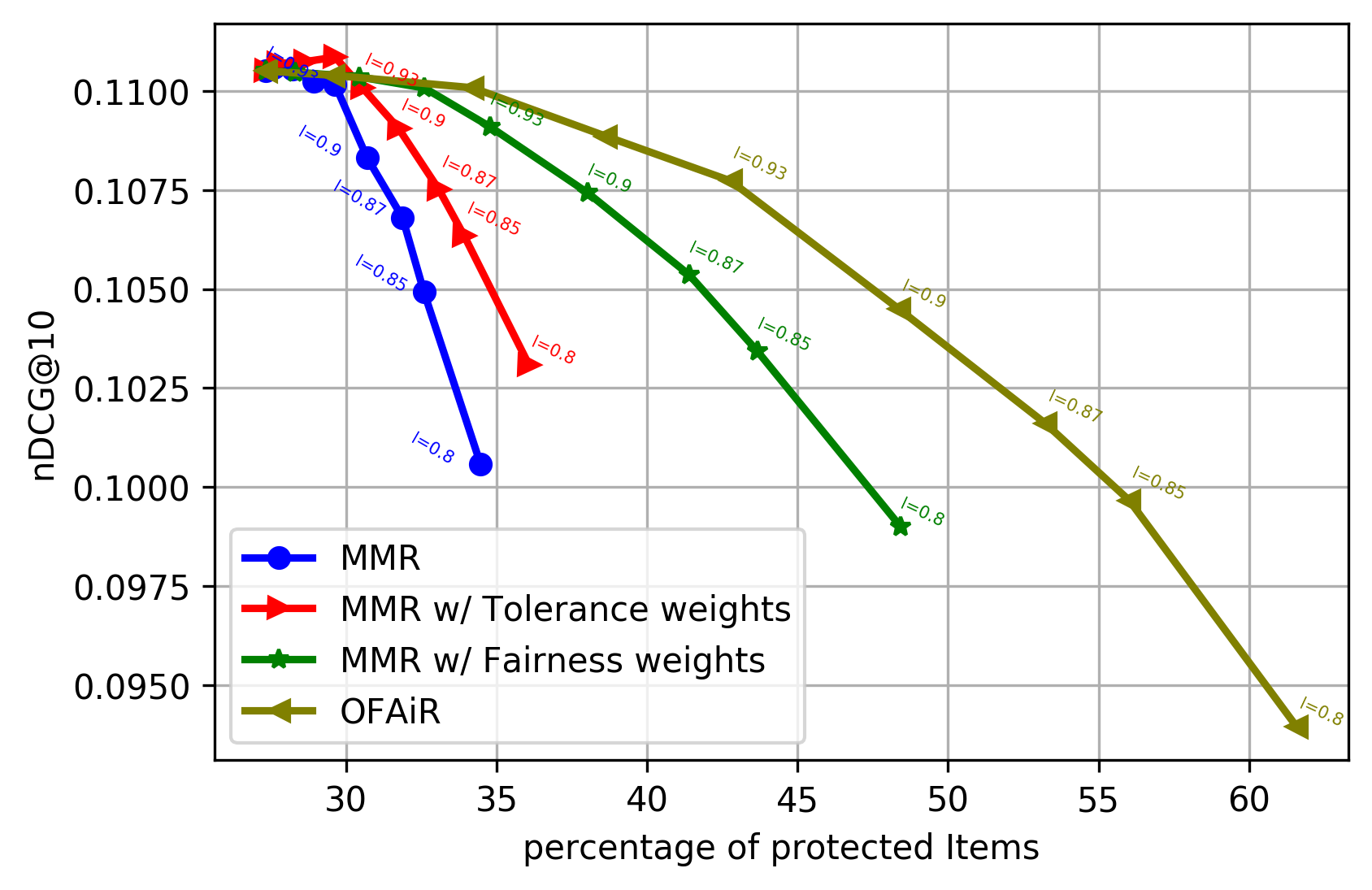}
    \caption{MMR-based re-ranking methods. The Movies Dataset.}
    \label{fig:ML_mmr_based_methods}
\end{figure}

\begin{figure}[tbh]
    \centering
    \includegraphics[width=\linewidth]{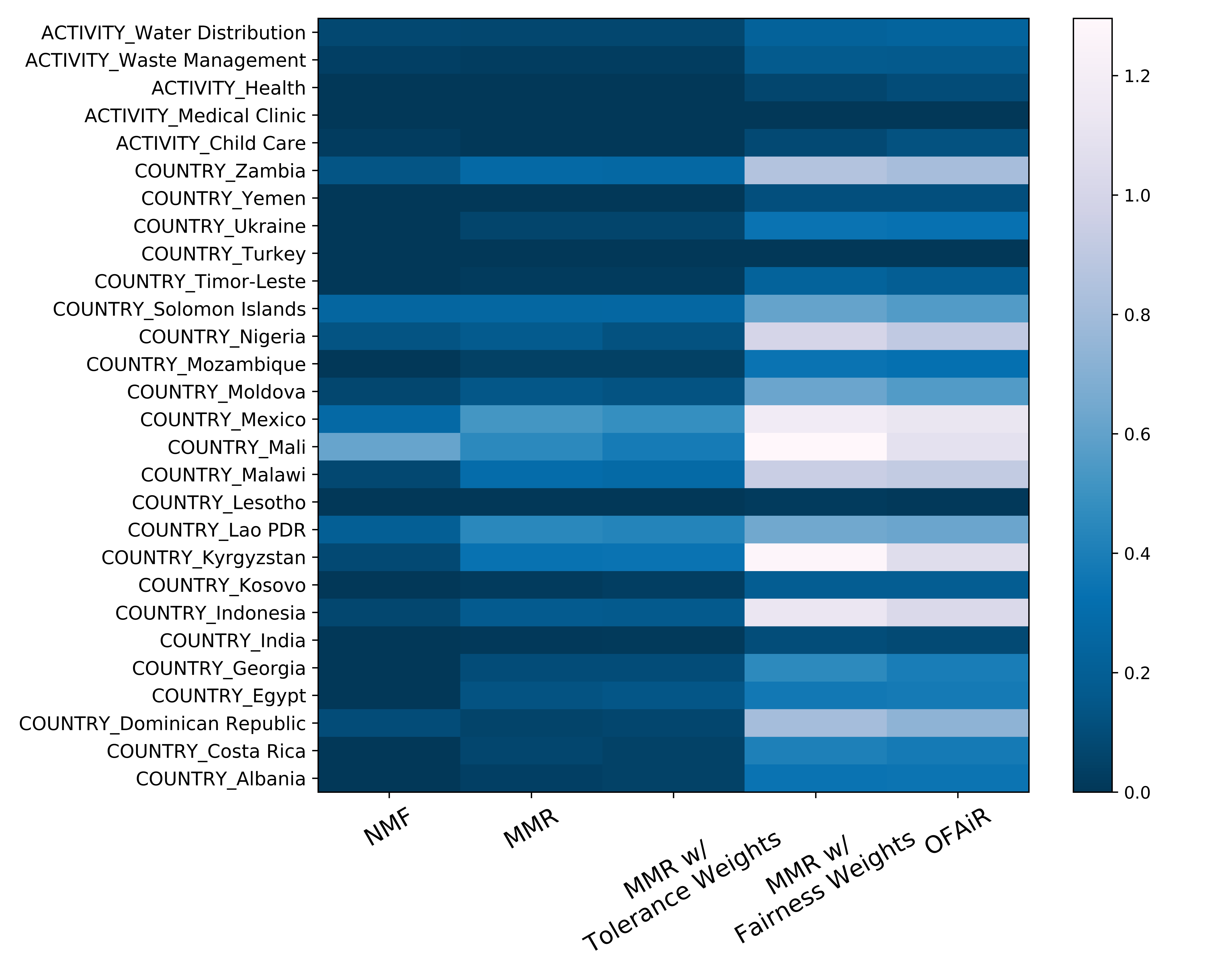}
    \caption{Cross-category fairness of MMR-based algorithms. Kiva dataset.}
    \label{fig:kiva_category_heatmap}
\end{figure}

\begin{figure}[tbh]
    \centering
    \includegraphics[width=\linewidth]{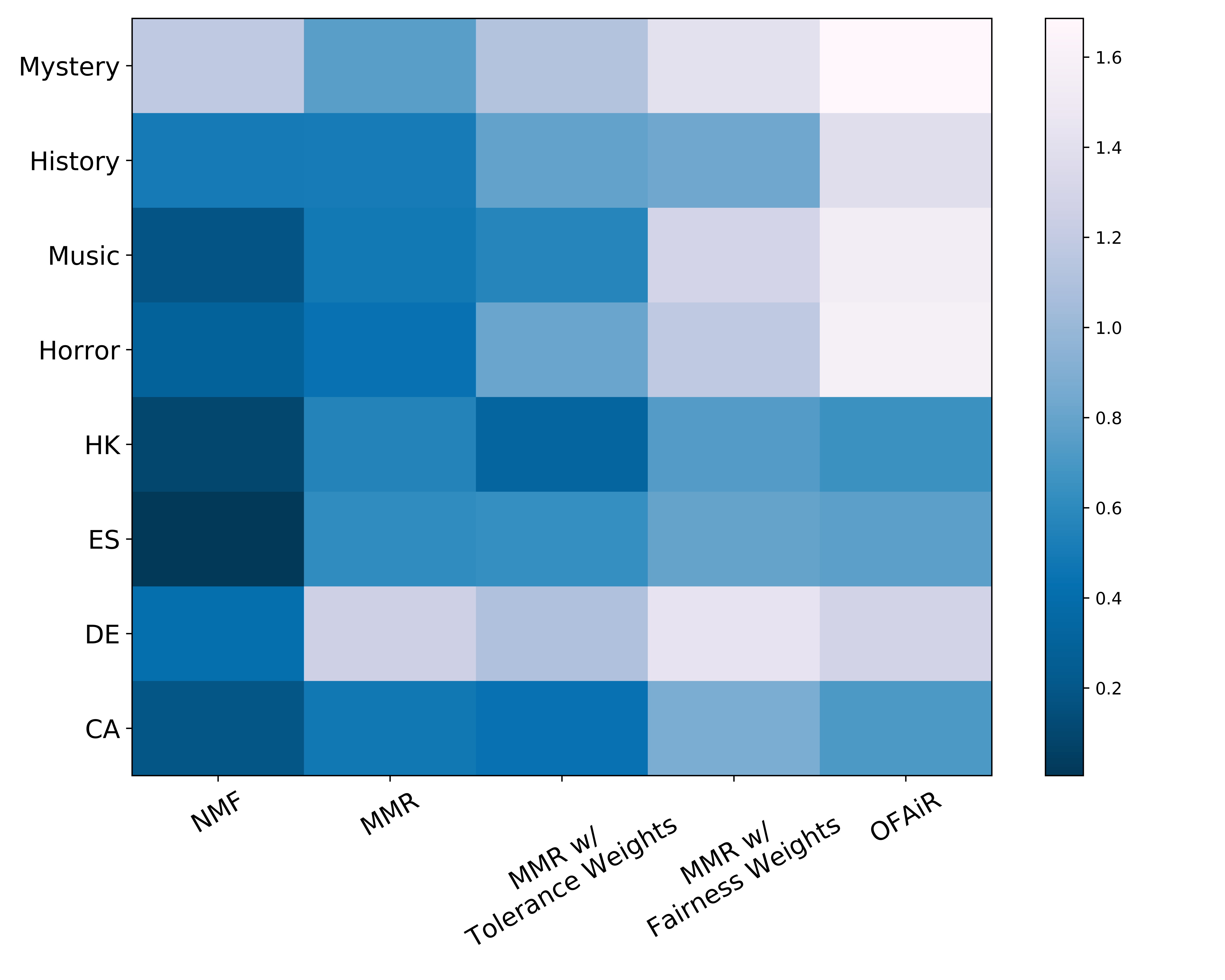}
    \caption{Cross-category fairness of MMR-based algorithms. The Movies dataset.}
    \label{fig:ML_category_heatmap}
\end{figure}

It is significant that OFAiR has a dominant position among the other algorithms in terms of the fairness / accuracy tradeoff when viewed across all items in the protected group.
However, a key objective of this work was to ensure distribution of fairness enhancement across multiple categories of protected groups. Figure \ref{fig:kiva_category_heatmap} and Figure \ref{fig:ML_category_heatmap} show this aspect of our experimental results.


In Figure \ref{fig:kiva_category_heatmap} and \ref{fig:ML_category_heatmap}, we can see the performance of all the algorithms in terms of improvement in the exposure of the protected items in each protected dimension in a more fined-grained manner. Recall that in the Kiva dataset, country and economic sector (shown as activity) were the sensitive features with 23 countries and 5 sectors labeled as protected. It is also worth mentioning that in both of these features, users had a high general entropy as well. The lighter colors show an improvement in fairness. As it is shown, the colors are darker in NMF and MMR. The right side of the heat-map contains lighter colors indicating more inclusion of protected items in recommendation lists. Lightest colors might belong to MMR with fairness weights, and after we add the tolerance weights to the algorithm it becomes slightly darker. This is due to the fairness/accuracy tradeoff noted above. For some feature values in \ref{fig:kiva_category_heatmap}, fairness is not improved by any algorithm. This is because the reranker can only improve the fairness of the results if these dimensions are present in the recommendation list of users and in these cases they rarely are. A similar trend is found in the Movies dataset, with the OFAiR algorithm, showing the best exposure across all of the protected dimensions.

\section{Related work}
In examining prior work on re-ranking, it is important to note the distinction introduced in Section~\ref{sec:background} above between user-oriented results diversification and fairness/provider-oriented re-ranking, which is the objective of our work. A user-oriented method will measure success by the diversity of individual lists, whereas a provider fairness approach will be measuring outcomes for providers, especially protected ones.

One of the first efforts to increase diversity in recommendations was \cite{Ziegler:2005:IRL:1060745.1060754}, which used a taxonomic content-based similarity metric to re-rank recommendation lists. This method did not attempt to personalize its ranking goal relative to different users. The taxonomic item similarity measure used in this work may be appropriate to adapt to OFAiR, which currently uses a one-dimensional representation of item features. A steady stream of user-oriented diversification research followed, as summarized in \cite{kunaver2017diversity}.

More closely related to the present work are the FAR/PFAR algorithms in \cite{liu2018personalizing,liu2019personalized}, which have served as an inspiration here. PFAR incorporates the individualized entropy-based user tolerance weight, thus enabling it to increase accuracy for the users with more fixed tastes. As noted above, however, PFAR is based on the aspect-oriented xQuAD algorithm, which has a binary inclusion objective. Once a provider is represented in the recommendation list, it is no longer boosted in re-ranking. This makes sense for the FAR/PFAR use case, which concentrates on fairness across multiple providers. This is less appropriate for a protected/unprotected binary distinction because the objective is satisfied with only a single protected item included and there is therefore no way to approach parity of representation. This can be seen in the very small improvements in exposure found with these algorithms. 

Another approach to fair ranking is the FA*IR algorithm proposed in \cite{zehlike2017fa}. This algorithm creates two queues: one of protected and one of unprotected items, and then integrates them to satisfy (in expectation) a probabilistic ranked fairness test. This algorithm does make the protected/unprotected assumption that we are using in this work. However, it applies only to a single such distinction. It might be interesting to extend the FA*IR model to multiple dimensions of fairness.

Fairness for multiple groups has been addressed in classification settings under the idea of \textit{rich sub-group fairness}~\cite{kearns2017preventing,kearns2019empirical}. In this work, the emphasis is on extending fairness guarantees to all possible combinations of protected groups in a dataset. The SUBGROUP algorithm alternately optimizes for a particular group's fairness and then seeks the group for whom fairness is most violated. In recommendation, we are not seeking a single decision rule, so we have a different solution in OFAiR: to distribute the optimization ``cost'' across different users in a personalized way.

\section{Conclusion and Future Work}\label{sect:conclusion}
The results of our experiments show that OFAiR works as intended. Its proportion-based MMR model provides a much better tradeoff between ranking accuracy and fairness for the protected-unprotected case than the FAR/PFAR models explored in prior work. In the datasets under study, we show that users' tolerance for diversity varies across features, which justifies our approach of differentiating users based on the opportunities they represent for enhancing provider-side fairness. 

We show that the combination of personalized, feature-specific, weights together with weights identifying protected feature values is effective with the feature-specific tolerance helping maintain accuracy and the feature weight promoting protected group items. As we showed, our method can be applied across multiple protected groups at the same time and can ensure fairness with respect to system's designed fairness goal for each feature.

One of the challenges in this work is the lack of proper datasets that have user features and these datasets are specifically lacking in domains where fairness matters. Due to this issue, we chose the Movies dataset to show the capabilities of our method.

In our future work, we intend to explore further the idea of ``opportunity'' in subgroup-fairness-aware recommendation. In particular, when recommendations are delivered over time, prior outcomes relative to different protected groups may dictate what opportunities should be most salient at any given moment.


\section{Acknowledgements}
This work was supported by the National Science Foundation under Grant No. 1911025.

\bibliographystyle{ACM-Reference-Format}
\balance
\bibliography{main.bib}

\end{document}